\documentclass{PoS}
    \usepackage[cyr]{aeguill}
\usepackage{subfigure}
\def\pT{\ensuremath{p_{\mathrm{T}}}} % Subscript roman not italic (EE)
\def\ET{\ensuremath{E_{\mathrm{T}}}} % Subscript roman not italic (EE)
\def\MEFF{\ensuremath{M_{\mathrm{eff}}}} % Subscript roman not italic (EE)
\def\MT{\ensuremath{M_{\mathrm{T}}}} % Subscript roman not italic (EE)
\def\aT{\ensuremath{\alpha_{\mathrm{T}}}} % Subscript roman not italic (EE)

\def\TeV{\ifmmode {\mathrm{\ Te\kern -0.1em V}}\else
                   \textrm{Te\kern -0.1em V}\fi}%
\def\GeV{\ifmmode {\mathrm{\ Ge\kern -0.1em V}}\else
                   \textrm{Ge\kern -0.1em V}\fi}%

\title{Prospects for R-Parity Conserving SUSY searches at the LHC}

\ShortTitle{Prospects for R-Parity Conserving SUSY searches at the LHC}

\author{\speaker{Marie-H\'el\`ene GENEST}%
         \thanks{The author would like to acknowledge support by the DFG cluster of excellence "Origin and Structure of the Universe" (www.universe-cluster.de).}\\
        On behalf of the CMS and ATLAS collaborations \\
        Ludwig-Maximilians-Universit\"at M\"unchen\\
        Am Coulombwall 1, 85748 Garching, Germany\\
        E-mail: \email{marie-helene.genest@physik.uni-muenchen.de}}

\abstract{We review the current strategies to search for generic SUSY models with R-parity conservation in the ATLAS and CMS detectors at the LHC. The discovery reach in early data will be presented for the different search channels based on missing transverse momentum from undetected neutralinos and multiple jets. We will also describe the search for models of gauge-mediated supersymmetry breaking for which the NLSP is a neutralino decaying to a photon and a gravitino. Finally, we will present recent work on techniques used to reconstruct the decays of SUSY particles at the LHC in early data, based on the selection of final-state exclusive decay chains.}

\FullConference{European Physical Society Europhysics Conference on High Energy Physics\\
		 July 16-22, 2009\\
		 Krakow, Poland}

\begin{document}

\section{Introduction}
Supersymmetric particle production at the LHC potentially involves long decay chains containing high-$\pT$ jets, leptons (from chargino or heavy neutralino decays) and missing transverse energy ($\slash\kern-.6em\ET$) from the lightest sparticle (LSP), if it is stable as in R-parity conserving models. Different search channels have been developed in order to discriminate a potential SUSY signal from the main background sources ($t\bar{t}$, QCD and W/Z+jets events) and strategies have been devised to find the SUSY parameters if a discovery is made. %Given $m_{\tilde{g}}=400\GeV$, gluino pairs should be produced approximately 20000 more frequently at the LHC (for $\sqrt{s}=14\TeV$) than at the Tevatron, while $t\bar{t}$ production will only be enhanced by a factor $\approx100$. 

\section{The different search channels}
The analysis is usually divided in different search channels, among which are:
\begin{itemize}
\item{The 0-lepton channel: The jets + $\slash\kern-.6em\ET$ analysis is the least model-dependent SUSY search. It needs strong cuts to reduce the QCD background, while keeping them simple enough in order to achieve high sensitivity. If the QCD background and sources of fake $\slash\kern-.6em\ET$ can be kept under control, the main backgrounds are $t\bar{t}$, W and Z production.}
\item{The 1-lepton channel: Requiring one lepton in addition to jets and $\slash\kern-.6em\ET$ is a slightly less general search, but it increases the robustness against the QCD background. The main backgrounds are $t\bar{t}$ and W production.}
\item{The multi-lepton channels: Heavy neutralinos or charginos in the decay chain can lead to 2 or 3 leptons in the final state. The 2-lepton search is subdivided into two channels: one asking for opposite-sign electric charges (non-resonant excess but non-negligible standard-model background) and the other, for same-sign electric charges (low standard-model background but lower statistics). The 3-lepton search also has two channels: one asking for an additional jet (no cut on $\slash\kern-.6em\ET$) and the other, for a large $\slash\kern-.6em\ET$ and a pair of same-flavour leptons with opposite-sign electric charges (to probe the direct gaugino production).}
%\item{The $\tau$ channel: This channel is well-suited for the large $\tan{\beta}$ region of the parameter space, as this region favours light $3^{rd}$ generation superpartners. The analysis requires the presence of hadronic taus; the leptonically decaying $\tau$ being indistinguishable from a lepton, they are already included in the leptonic channels. The main backgrounds are the same as for the 0-lepton channel.}
%\item{The b channel: The presence of b-jets in SUSY cascades is also enhanced by large $\tan{\beta}$; this analysis asks for 2 or 3 b-jets in the final state (the standard b-tagging in ATLAS has $\epsilon~50-60\%$). The main background is W+jets.}
\end{itemize}
There are also many other channels, e.g. channels based on $\tau$ or b jets. More details can be found in \cite{cscnote} and \cite{cmstdr}.

\section{Examples of mSUGRA analyses}

\subsection{The 1-lepton channel in ATLAS at $\sqrt{s}=10\TeV$}

As detailed in \cite{10tev}, this analysis requires one isolated lepton (electron or muon) with $\pT>20\GeV$ and no other lepton with $\pT>10\GeV$. Furthermore, at least 4 jets J$_\mathrm{i}$ (i=1-4, ordered in decreasing $\pT$) must be reconstructed with $\pT^{\mathrm{J}_\mathrm{2-4}}>40\GeV$ and $\pT^\mathrm{J_\mathrm{1}}>100\GeV$. To avoid fake $\slash\kern-.6em\ET$ from a mismeasured jet, $\Delta\phi(\slash\kern-.6em\ET,\mathrm{J}_\mathrm{i})$ must be greater than 0.2 for i=1-3. A large $\slash\kern-.6em\ET$ is required, $\slash\kern-.6em\ET>\max(80\GeV$$,0.2\MEFF)$, along with a transverse sphericity greater than 0.2 and a large transverse mass, $\MT>100\GeV$, where ${\MEFF=\pT^{\mathrm{lepton}}+\displaystyle\sum_{\mathrm{i}}^{}\pT^\mathrm{J_\mathrm{i}}+\slash\kern-.6em\ET}$ and $\MT=\sqrt{2\pT^{\mathrm{lepton}}\slash\kern-.6em\ET(1-\cos{\Delta\phi})}$.
Figure \ref{fig:meff} shows $\MEFF$ after all cuts for the ATLAS benchmark point SU4\cite{cscnote}; the signal clearly exceeds the background.

\subsection{Discovery reach}

Figure \ref{fig:reach} shows the ATLAS expected $5\sigma$ discovery reach for 200 pb$^{-1}$ of data at $\sqrt{s}=10\TeV$ in the mSUGRA $m_{1/2}$ versus $m_0$ plane ($\tan{\beta}=10$, $A_0=0$ and $\mu>0$), including systematic errors\cite{10tev}. From this Figure, one can see that the 0- and 1-lepton channels have the highest reach and that squarks and gluinos with masses up to 750$\GeV$ can be discovered. A similar discovery potential is obtained by the CMS collaboration.

\subsection{The 0-lepton dijet channel in CMS at $\sqrt{s}=14\TeV$}

\begin{figure}[htbp]
  \begin{center}
    \subfigure[The 1-lepton channel in ATLAS]{
        \label{fig:meff}
        \includegraphics[scale=0.34,angle=0]{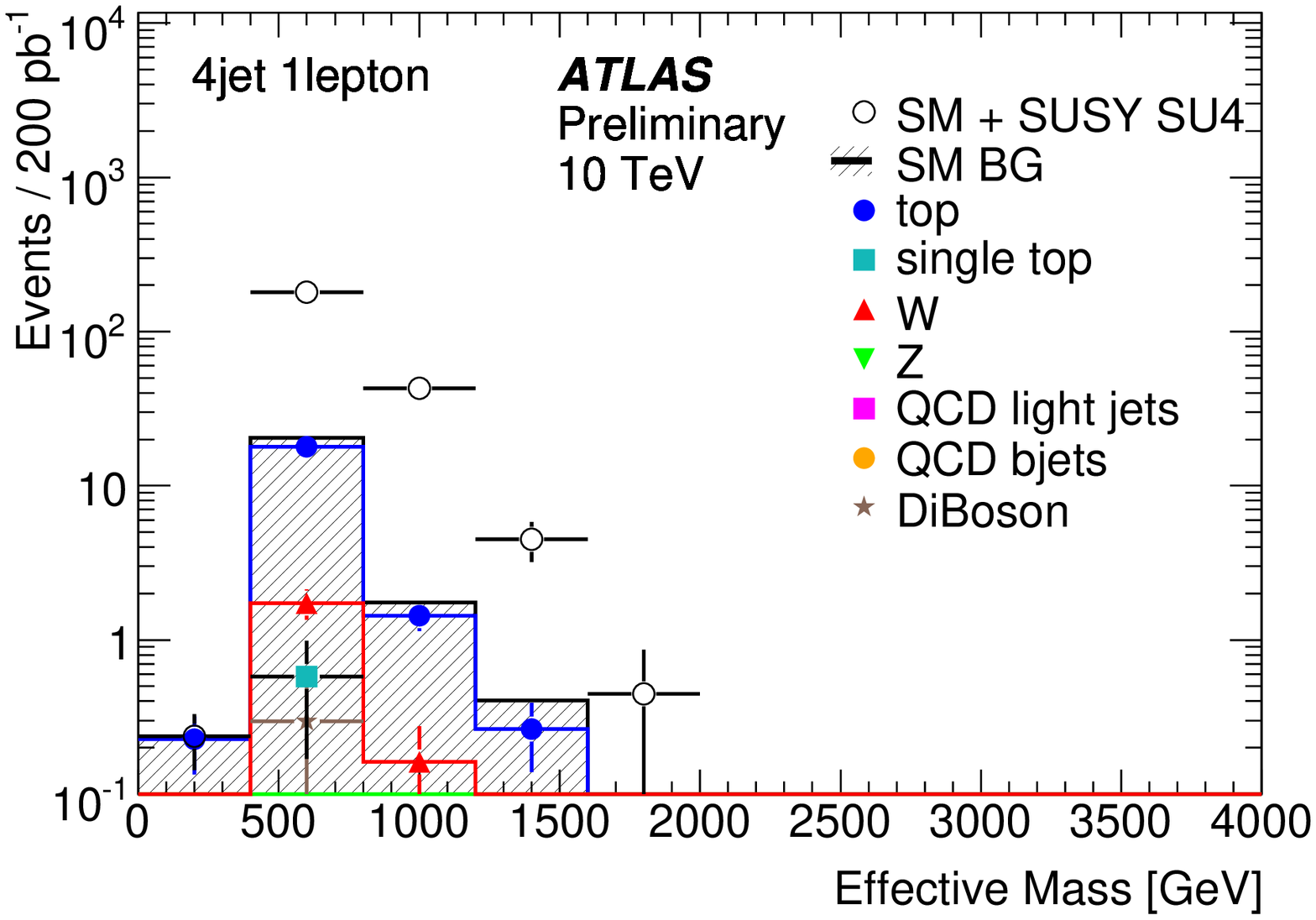}
    }
    \subfigure[Discovery potential for 200 pb$^{-1}$]{
        \label{fig:reach}
        \includegraphics[scale=0.18,angle=0]{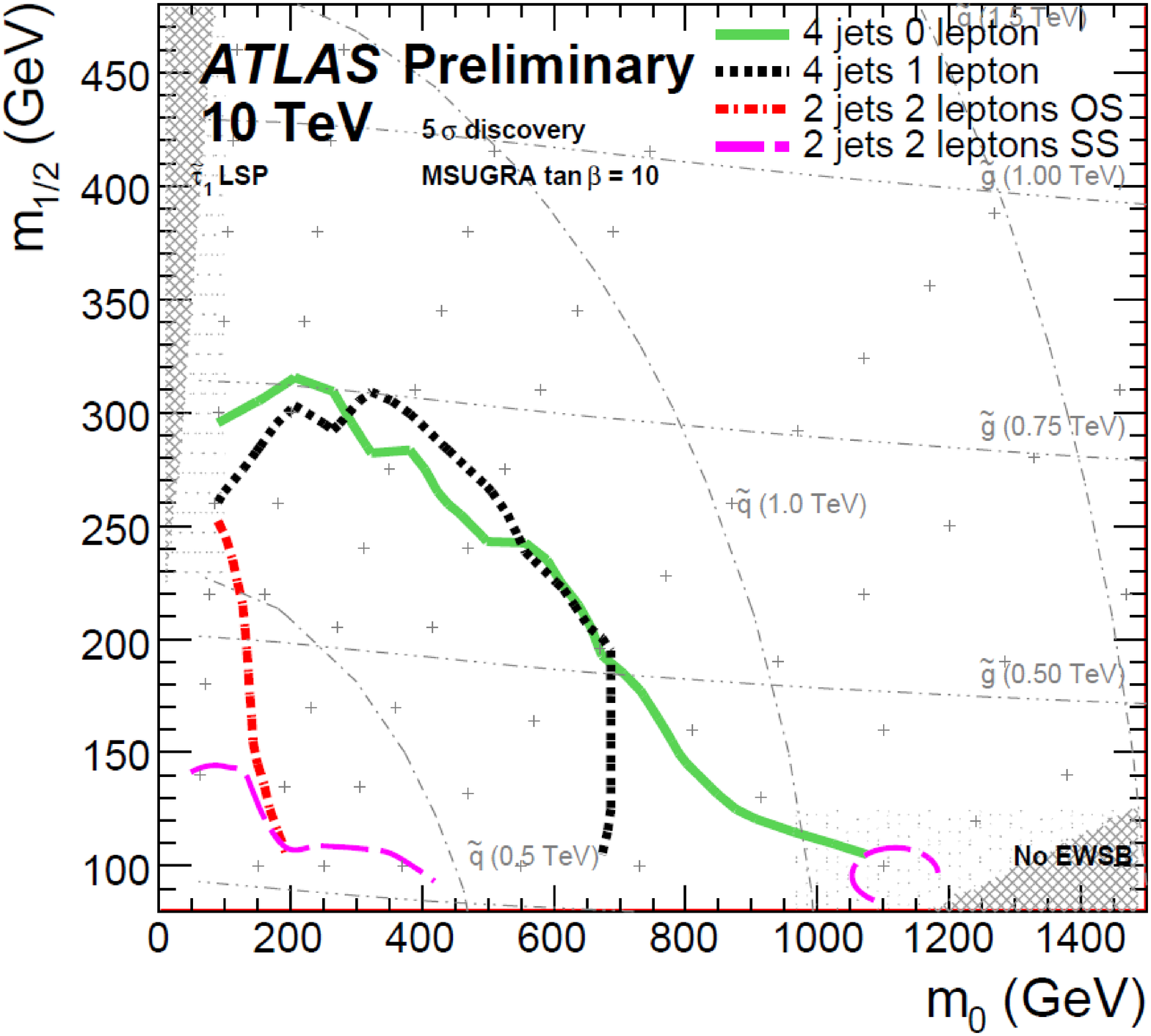}
    }
    \subfigure[The 0-lepton dijet channel in CMS]{
        \label{fig:alphaT}
        \includegraphics[scale=0.37,angle=0]{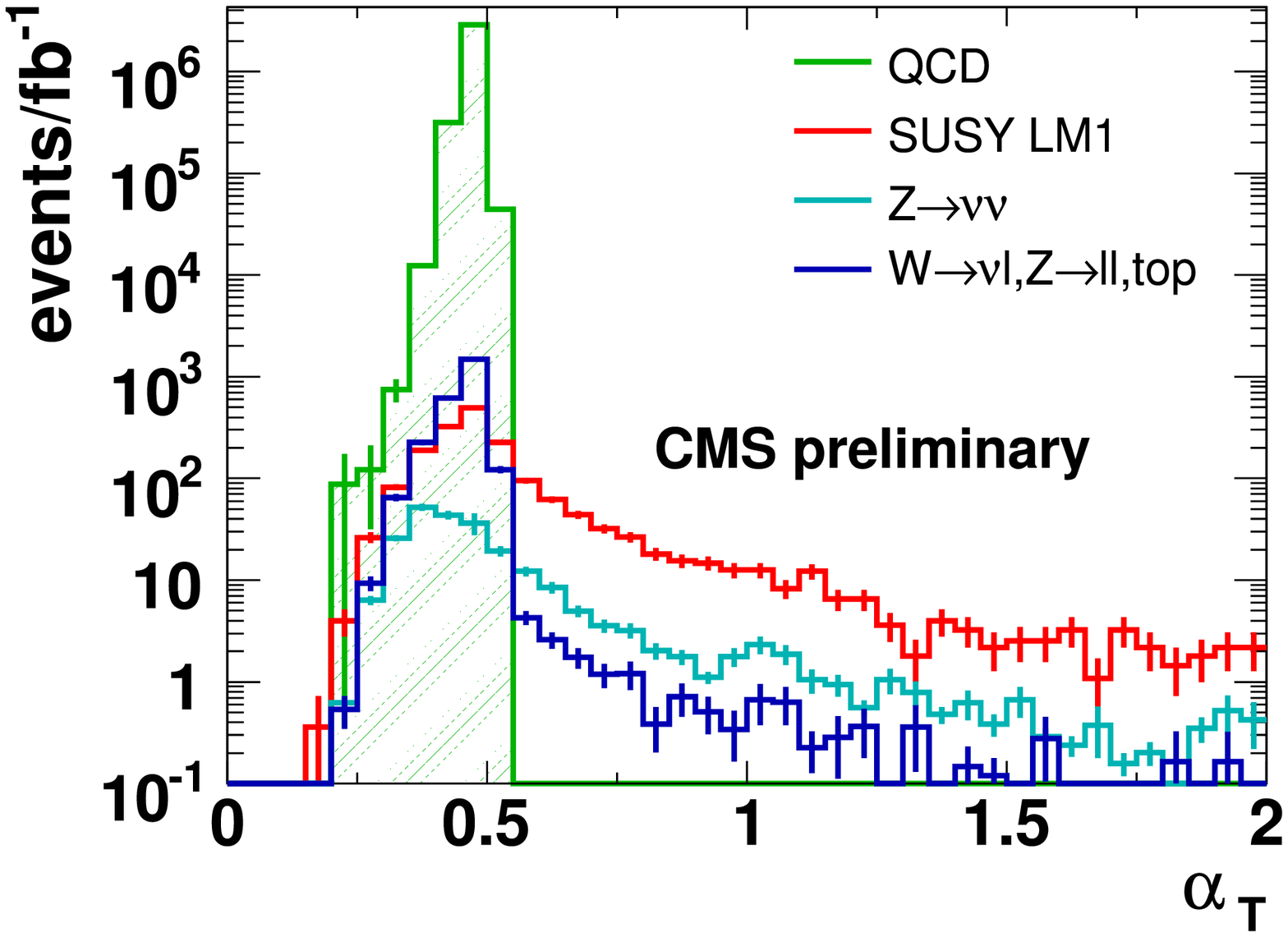}
    }
  %  \subfigure[GMSB analysis in ATLAS]{                                                                                                                                                                             \label{fig:gmsb}                                                                                                                                                                                   \includegraphics[scale=0.39,angle=0]{gmsb.eps}
   % }
    \subfigure[Exclusive measurement in CMS]{
    	\label{fig:mmumu}
   	\includegraphics[scale=0.38,angle=0]{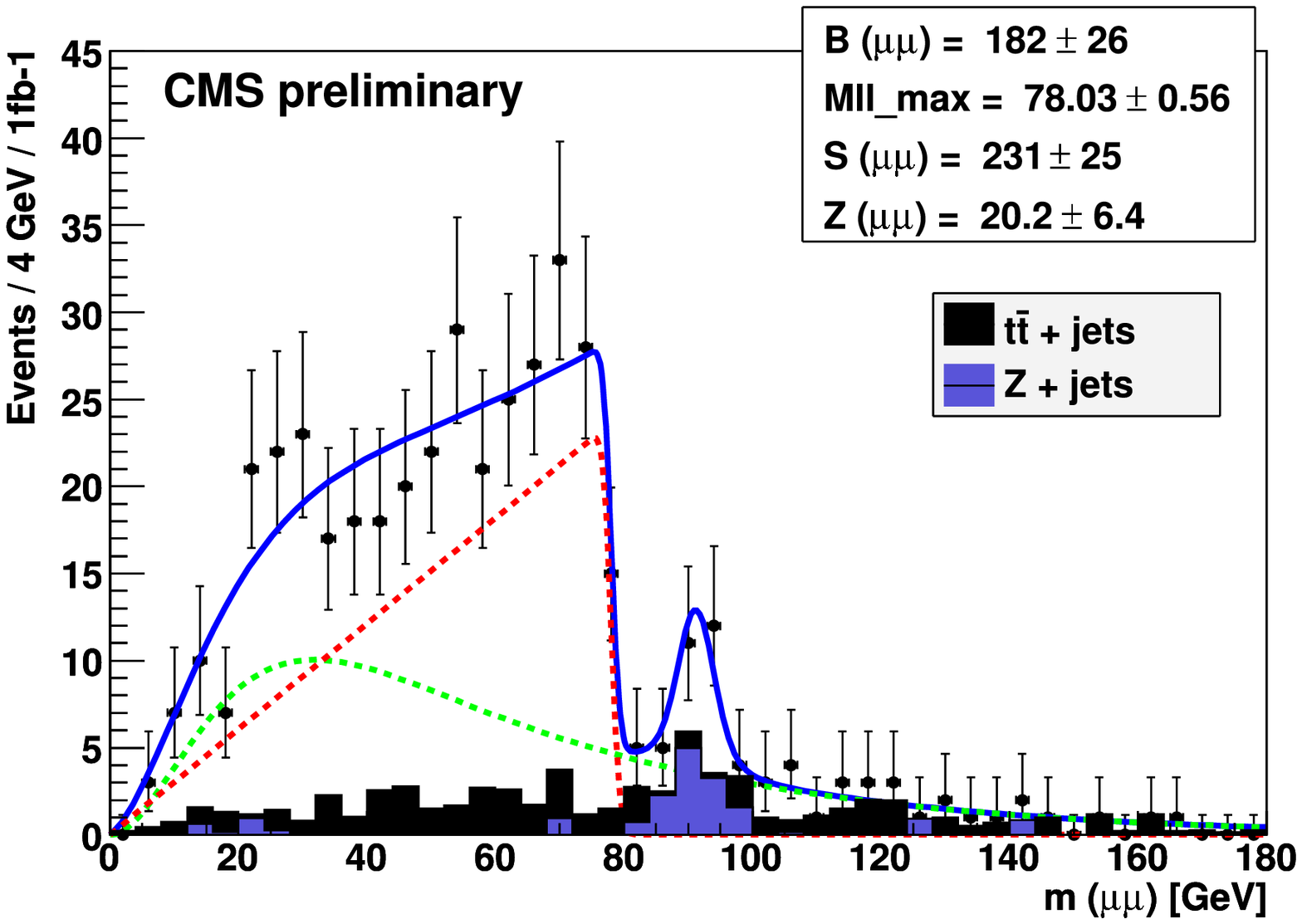}
    }
  \end{center}
  \caption{ (a) $\MEFF$ distribution after all the cuts for the benchmark point SU4 ($m_0$=200$\GeV$, $m_{1/2}$=160$\GeV$, $A_0$=-400$\GeV$, $\tan{\beta}$=10 and $\mu>0$) and the backgrounds for 200 pb$^{-1}$ of data at $\sqrt{s}=10\TeV$ (b) Expected $5\sigma$ discovery potential for 200 pb$^{-1}$ of data at $\sqrt{s}=10\TeV$ (c) $\aT$ after all the cuts for the benchmark point LM1 ($m_0$=60$\GeV$, $m_{1/2}$=250$\GeV$, $A_0$=0, $tan{\beta}$=10 and $\mu>0$) and the backgrounds for 1 fb$^{-1}$ of data at $\sqrt{s}=14\TeV$ (d) The di-muon invariant mass $m_{\mu\mu}$ in CMS along with the fit (in blue), which includes the contributions from the flavour-symmetric background (in green) extracted from data, the
 signal (in red) and the Z mass peak. %Number of reconstructed photons with $\pT>20\GeV$ and $|\eta|<2.5$ after all cuts for the benchmark point GMSB1 ($\Lambda$=90 TeV, $M$=500 TeV, $C_{grav}$=1.0, $N$=1, $\tan{\beta}$=5, $\mu>0$) and the backgrounds for 1 fb$^{-1}$ of data at $\sqrt{s}=14\TeV$.
  \label{fig1}}
\end{figure}

This analysis, inspired by \cite{randall}, aims at probing the $\tilde{q}\rightarrow q \tilde{\chi}_1^0$ channel ($m_{\tilde{q}}<m_{\tilde{g}}$), which has a 2-jet + $\slash\kern-.6em\ET$ signature. The signal can be discriminated from the QCD background by using the variable $\aT= \ET^{\mathrm{J}_\mathrm{2}}/\sqrt{2\ET^{\mathrm{J}_\mathrm{1}}\ET^{\mathrm{J}_\mathrm{2}}(1-\cos{\Delta\phi})}$, which would give exactly 0.5 for a perfectly measured QCD event. As detailed in \cite{dijet}, the analysis asks for two jets with $\pT>50\GeV$, the most energetic one also satisfying $|\eta|<2.5$. Furthermore, the transverse momentum of the jets must be so that their scalar sum exceeds 500$\GeV$ and that ${\Delta\phi(-|\displaystyle\sum_{\mathrm{jets}}^{}\vec{p}_\mathrm{T}|,\mathrm{J}_\mathrm{i})>0.3}$, where i=1,2,3. Finally, the events are rejected if they contain a third jet with $\pT>50\GeV$ or a lepton (electron or muon) with $\pT>10\GeV$. Figure \ref{fig:alphaT} shows $\aT$ after all these cuts: cutting on $\aT>0.55$ removes all the remaining QCD background. For 1~fb$^{-1}$ of data at ${\sqrt{s}=14\TeV}$, one expects, after this final cut, 439 SUSY events against 77 background events: a discovery can be made without using the full $\slash\kern-.6em\ET$ calculation for the CMS benchmark point LM1\cite{dijet}.

\section{Gauge mediated supersymmetry breaking}

One must also consider SUSY models which predict different signatures, like gauge mediated supersymmetry breaking (GMSB) for which the LSP is the gravitino. In this model, the NSLP can be the lightest neutralino which would decay into a photon and a gravitino, leading to a signature containing jets, $\slash\kern-.6em\ET$ and photons. Such a study has been made in ATLAS\cite{cscnote}. If the decay length of the neutralino is comparable to the size of the ATLAS inner detector, there could be high-$\pT$ photons not originating from the interaction point ("non-pointing photons"), which can have a wider shower profile than pointing photons. In order to identify these photons, one must loosen the photon identification cuts to keep only the ones which are unbiased with respect to the neutralino decay length. For example, the width of the cluster in the $2^{\mathrm{nd}}$ layer of the electromagnetic calorimeter, a variable normally used in photon identification, depends on the origin of the photon emission and must therefore be dropped from the non-pointing photon identification. %After standard jets and $\slash\kern-.6em\ET$ selection cuts ($N_{jets}\ge4$ with $\pT>50\GeV$ and $\pT^{J1}>100\GeV$, $\slash\kern-.6em\ET>$max$(100\GeV$$,0.2\MEFF)$), the number of loose-identification photons can serve as a good discriminant against the main background ($t\bar{t}$), as can be seen in Figure \ref{fig:gmsb}. 
By removing the biased cuts, the non-pointing photon identification efficiency changes from $36.1\pm0.6\%$ to $80.7\pm0.5\%$ for the ATLAS benchmark point GMSB3 ($\Lambda=90\TeV$, $M_\mathrm{m}=500\TeV$, $C_\mathrm{G}=55.0$, $N_\mathrm{5}=1$, $\tan{\beta}=5$, $\mu>0$). Doing so, the fraction of jets reconstructed as photons increases from $0.19\pm0.03\%$ to $0.70\pm0.07\%$.  Techniques are also being developed to extract the lifetime of the neutralino using timing and directional information from the liquid argon calorimeter system.

\section{Exclusive measurements}
If SUSY is discovered, the next step will be to measure the sparticle mass spectrum and derive the parameters of the model using the kinematics of long decay chains, like ${\tilde{g}\rightarrow q\tilde{q}\rightarrow q\tilde{\chi}_2^0 \rightarrow l\tilde{l}\rightarrow l\tilde{\chi}_1^0}$. Since the LSP is not detected, the decay chain cannot be completely reconstructed: edges, rather than mass peaks, are measured in the invariant mass of the decay products (e.g. $m_{ll}$, $m_{llq}$, $m_{lq}$, etc). For example, Figure~\ref{fig:mmumu} shows $m_{\mu\mu}$ as would be measured by CMS for the decay $\tilde{\chi}_2^0 \rightarrow ll\tilde{\chi}_1^0$ of the benchmark point LM1 \cite{cmsdilep}. The end-point $m_{\mu\mu}^{\mathrm{max}}$ is related to the sparticle masses by the formula ${m_{ll}^{\mathrm{max}}=m_{\tilde{\chi}_2^0}\sqrt{1-m_{\tilde{l}}^2/m_{\tilde{\chi}_2^0}^2}\sqrt{1-m_{\tilde{\chi}_1^0}^2/m_{\tilde{l}}^2}}$. To extract this end-point, $m_{\mu\mu}$ is fitted with the function ${F(m)=N_{\mathrm{s}}S(m)+N_{\mathrm{bg}}B(m)+N_\mathrm{Z}Z(m)}$, where $S(m)$ is the signal model, a convolution of the theoretical shape with a resolution model, $B(m)$ describes the flavour-symmetric background (e.g. $t\bar{t}$, $WW$ or other SUSY cascades) which is taken from $e\mu$ data (as these backgrounds should produce $\mu\mu$ and $e\mu$ with equal probability), Z(m) models the Z peak, and $N_{\mathrm{s}}$, $N_{\mathrm{bg}}$ and $N_\mathrm{Z}$ are the number of events corresponding to each of these cases.

 Measuring the different end-points allows one to extract the masses of the sparticles even with only 1 fb$^{-1}$ of data at $\sqrt{s}=14\TeV$\cite{cscnote}. For example, $m_{\tilde{q}}=614\pm91\pm11\GeV$ can be extracted from simulated data with $m_{\tilde{q}}^{\mathrm{MC}}=634\GeV$. In a similar way, one can extract $m_{\tilde{\chi}_1^0}$, $m_{\tilde{\chi}_2^0}$ and $m_{\tilde{l}}$ with relatively large errors due to the limited statistics; the mass differences can however be measured with greater accuracy, e.g. $m_{\tilde{\chi}_2^0}-m_{\tilde{\chi}_1^0}=100.6\pm1.9\pm0.0\GeV$ ($\Delta m^{\mathrm{MC}}=100.7\GeV$). The ultimate goal remains to determine the SUSY parameters and this can be achieved by feeding the end-point measurements to the Fittino package; studies on simulated data show that for 1 fb$^{-1}$ of data at $\sqrt{s}=14\TeV$, $m_0$ and $m_{1/2}$ can be well constrained (e.g. $m_0^{\mathrm{fit}}=98.5\pm9.3\GeV$ for $m_0^{\mathrm{MC}}=100\GeV$) and the right sign of $\mu$ favoured, while $A_0$ and $\tan{\beta}$ are more problematic since there is no information from the Higgs sector at low integrated luminosity. The CMS collaboration has similar studies and findings.
%\begin{figure}[htbp]
%  \begin{center}
%    \includegraphics[scale=0.3,angle=0]{m_mumu.eps}
%  \end{center}
%  \caption{ The di-muon invariant mass $m_{\mu\mu}$ in CMS along with the fit (in blue), which includes the contributions from the flavour symmetric background (in green) extracted from data, the signal (in red) and the Z mass peak.
%  \label{fig:mmumu}}
%\end{figure}

\section{Conclusions}
Looking for excesses in inclusive search channels covering different scenarios, like mSUGRA and GMSB, is the first step towards finding SUSY at the LHC. The LHC should be able to discover squarks and gluinos with masses up to 750$\GeV$ with 200 pb$^{-1}$ of data at $\sqrt{s}=10\TeV$. After a discovery, the next step would be to select specific decay chains to measure the properties of the new particles. Some masses and parameters could already be extracted with 1 fb$^{-1}$ of data at $\sqrt{s}=14\TeV$ for low-mass scenarios.

\end{document}